\documentclass[journal]{IEEEtran}

\ifCLASSINFOpdf
\else
   \usepackage[dvips]{graphicx}
\fi
\usepackage{url}

\usepackage{graphicx}
\usepackage{amsmath}

\begin{document}

\title{Phase-Center-Constrained Beamforming for Minimizing Phase-Center Displacement}

\author{
    \IEEEauthorblockN{
        Jan Steckel\IEEEauthorrefmark{1}\IEEEauthorrefmark{2}\IEEEauthorrefmark{5},
        Noori Bni-Lam \IEEEauthorrefmark{4} 
    }\\
    \IEEEauthorblockA{\IEEEauthorrefmark{1}Cosys-Lab, Faculty of Applied Engineering, University of Antwerp, Antwerp, Belgium\\}
    \IEEEauthorblockA{\IEEEauthorrefmark{2}Flanders Make Strategic Research Centre, Lommel, Belgium\\}
    \IEEEauthorblockA{\IEEEauthorrefmark{4}ESA/ESTEC, Keplerlaan 1, 2201 AZ Noordwijk, the Netherlands.\\}
    \IEEEauthorblockA{\IEEEauthorrefmark{5}jan.steckel@uantwerpen.be}
 }

\markboth{  }
{Steckel \MakeLowercase{\textit{et al.}}: Phase-Center-Constrained Beamforming for Minimizing Phase-Center Displacement}
\maketitle

\begin{abstract}
Accurate knowledge and control of the phase center in antenna arrays is essential for high-precision applications such as Global Navigation Satellite Systems (GNSS), where even small displacements can introduce significant localization errors. Traditional beamforming techniques applied to array antennas often neglect the variation of the phase center, resulting in unwanted spatial shifts, and in consequence, localization errors. In this work, we propose a novel beamforming algorithm, called  Phase-Center-Constrained Beamforming (PCCB), which explicitly minimizes the displacement of the phase center (Phase Center Offset, PCO) while preserving a chosen directional gain. We formulate the problem as a constrained optimization problem and incorporate regularization terms that enforce energy compactness and beampattern fidelity. The resulting PCCB approach allows for directional gain control and interference nulling while significantly reducing PCO displacement. Experimental validation using a simulated GNSS antenna array demonstrates that our PCCB approach achieves a fivefold reduction in PCO shift compared to the PCO shifts obtained when using conventional beamforming. A stability analysis across multiple random initializations confirms the robustness of our method and highlights the benefit of repeated optimization. These results indicate that our PCCB approach can serve as a practical and effective solution for decreasing phase center variability.
\end{abstract}


\IEEEpeerreviewmaketitle

\section{Introduction}
\IEEEPARstart{A}{ntenna} arrays are an important factor in many industrial communication applications. Through the coherent processing of multiple spatially diverse antennas, significant benefits can be brought to the communication fidelity. Indeed, various famous algorithms such as Bartlett beamforming and MVDR beamforming allow increasing the selectivity of the antenna system in order to suppress noise and interferences \cite{treesOptimumArrayProcessing2002}, and high-resolution angle-of-arrival estimation techniques have been developed such as the famous MUSIC algorithm  \cite{treesOptimumArrayProcessing2002}. Over the years, antenna arrays have found their way in a wide range of applications, such as low-power communication \cite{moto:c:irua:148627_bni_adap, moto:c:irua:156479_stec_hard, moto:c:irua:166458_bni_angl}, Radar \cite{haoBootstrappingAutonomousDriving2024} and telecommunication \cite{gharbiDesignPatchArray2017, zhang5GMillimeterWaveAntenna2017}.

When moving towards high-resolution GNSS applications (with localization accuracies in the centimeter or millimeter range), the location of the so-called phase-center of the receiving antenna plays a crucial role. Accurate knowledge of the so-called Phase Center Offset (PCO), and Phase Center Variation (PCV) is necessary when moving towards high-resolution GNSS implementations. The PCO is the location of a virtual point in the coordinate system of the antenna which best represents the spatial phase response of an point-like receiver \cite{chenDeterminingThreedimensionalPhase2014}, whereas the PCV is the virtual deviation of the PCO to explain the deviation of the phase response in a certain spatial direction.

Given the fact that the PCO and PCV of the receiving antenna are crucial aspects in high-resolution GNSS implementations, special care must be taken when designing beamforming algorithms for GNSS. Typically, this special care is taken through the implementation of rigorous calibration methods \cite{daneshmandPreciseCalibrationGNSS2014} which quantify the PCO and PCV under various beamforming conditions, that can subsequently be used to implement corrections on the pseudo-range measurements of the GNSS system \cite{borowskiPhaseCentreCorrections2022}. This calibration procedure, however, is a tedious process requiring anechoic chamber measurements, which cannot always be carried out in practice. In this paper, we therefore propose a novel beamforming algorithm, called Phase-Center-Constrained beamforming, which adjusts the set of weights of the beamformer to minimize the PCO and PCV of the antenna array under arbitrary beamforming conditions.

The rest of the paper is structured as follows. First, we will introduce the signal model and notation of the array antenna. Next, we derive the PCCB algorithm to calculate the weightvector that minimizes the PCO/PCV, after which we provide experimental validation of the approach. We conclude the paper with  a discussion of the results, and point towards future directions for improvement of the PCCB method.

\section{Introuction of Signal model}
Without loss of generality, we assume a planar sensor array of N elements, located in the YZ plane of the coordinate system of the problem (see figure \ref{fig:coord}). From direction $\psi$, a plane wave source with wavelength $\lambda$ originates and impinges onto the array. We can define the azimuth angle $\theta$ and elevation angle $\varphi$ as shown in figure \ref{fig:coord}. The coordinates of the sensors are in the form $\begin{bmatrix} 0 & p_y & p_z \\ \end{bmatrix}^T$. As is typical in array signal processing, we calculate a steering vector of the array \cite{treesOptimumArrayProcessing2002}, steered in direction $\psi$ as follows:

\begin{figure}[t]
    \centering
    \includegraphics[width=0.8\linewidth]{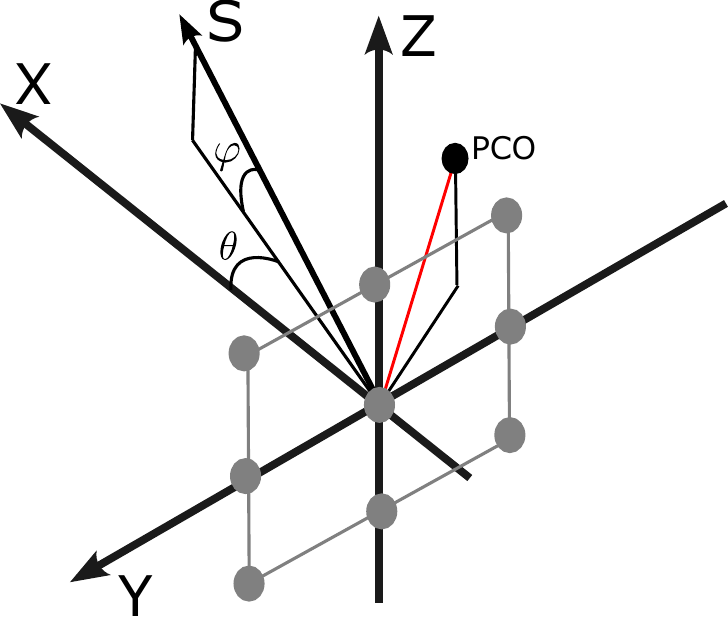}
    \caption{Coordinate system of the planar antenna array used in the PCCB formulation. The array is positioned in the YZ-plane, with incoming plane waves defined by azimuth angle $\theta$ and elevation angle $\varphi$. Sensor positions are denoted by grey circles, and the wave propagation direction is indicated by the vector S. The PCO is indicated by a black circle, and the PCO norm by a red line.} 
    \label{fig:coord}
\end{figure}

\begin{equation}
    \vec{v}(\psi) = e^{\frac{2 \cdot \pi \cdot i}{\lambda} (\psi_Y + \psi_Z) }
\end{equation}
where $\vec{v}(\psi)$ has dimensions $[N \times 1]$ (one complex value for each sensor element), and where the position and direction dependent components $\psi_Y$ and $\psi_Z$ are calculated as:
\begin{equation}
\begin{aligned}
    \psi_y(n) &= \sin(\theta)\cdot \cos(\varphi)\cdot p_y(n) \\
    \psi_z(n) &= \sin(\varphi)\cdot p_z(n)
\end{aligned}
\end{equation}
for the n-th sensor element. We can combine all the steering vectors in the steering matrix $V$ as follows:
\begin{equation}
 V = \begin{bmatrix}
\vec{v}(\psi_1) & \vec{v}(\psi_2)  & \cdots & \vec{u}(\psi_N) \\
\end{bmatrix}
\end{equation}
This steering matrix has dimensions $[N \times K]$ where $K$ is the number of sampled directions. Typically, this spatial sampling is performed using a uniform sampling on the sphere with the constraint of constant solid angle \cite{reijniersOptimizedSpatialSampling2020}. We can then define the unit-norm direction vectors $\vec{u}(\psi)$ which point in direction $\psi$. We are now ready to introduce the PCCB algorithm.

\section{Derivation of the PCCB algorithm}

Let us consider a desired "look direction" $\psi$, from which we want to receive a signal. We then can define the location of the Phase Center Offset as follows \cite{kunyszAntennaPhaseCenter2010a, liOptimizedLeastSquare2023}:
\begin{equation}
\vec{P}_{\text{c}} \cdot \frac{\vec{u}(\psi)}{\|\vec{u}(\psi)\|} = \frac{\lambda}{2\pi}\Delta\varphi(\psi)    
\end{equation}
where $\Delta\varphi(\psi) $ is the phase angle of the complex directivity pattern of the array antenna in direction $\psi$. We continue with defining a direction matrix $D$ and phase vector $\Phi$ as:

\begin{equation}
D = \begin{bmatrix}
\vec{u}(\psi_1)^T \\
\vdots \\
\vec{u}(\psi_K)^T
\end{bmatrix}_{K \times 3}, \quad
\Phi = \begin{bmatrix}
\Delta \varphi(\psi_1) \\
\vdots \\
\Delta \varphi(\psi_K)
\end{bmatrix}_{K \times 1}
\end{equation}
with matrix sizes $[K \times 3 ]$ for matrix $D$ and $[K \times 1 ]$ for vector $\Psi$. We can express the phase center $\vec{P}_c$ using $D$ and $\Psi$ as follows:
\begin{equation}
D \cdot \vec{P}_c = \frac{\lambda}{2\pi}\Phi \quad \Rightarrow \quad \vec{P}_c = \frac{\lambda}{2\pi}D^{-1}\Phi
\label{eq:PCO}
\end{equation}
Now, consider a beamformer with weights $w$, which gives the complex directivity pattern $B(\psi)$ as:
\begin{equation}
B_w(\psi) = w^H \cdot V
\end{equation}
The goal of the PCCB algorithm is to find the optimal $w$ which minimizes the PCO, while still maintaining a main-lobe in $B(\psi)$ in the desired look direction $\psi_D$. When analyzing the formulation, we see that $\Psi$ is dependent on $w$, as it depends on the complex directivity pattern $B(\psi)$. We define the norm of the PCO as $\gamma$:

\begin{equation}
\gamma = \|\vec{P}_c\|_2^2
\end{equation}
From here, we can more accurately define the phase differences $\Delta\varphi(\psi)$ as:
\begin{equation}
\Delta\varphi(\psi) = \angle B_w(\psi) = \angle w^H V
\end{equation}
where $\angle c$ stands for the angle-operator, extracting the phase from the complex number $c$.
\begin{figure}[t]
    \centering
    \includegraphics[width=\linewidth]{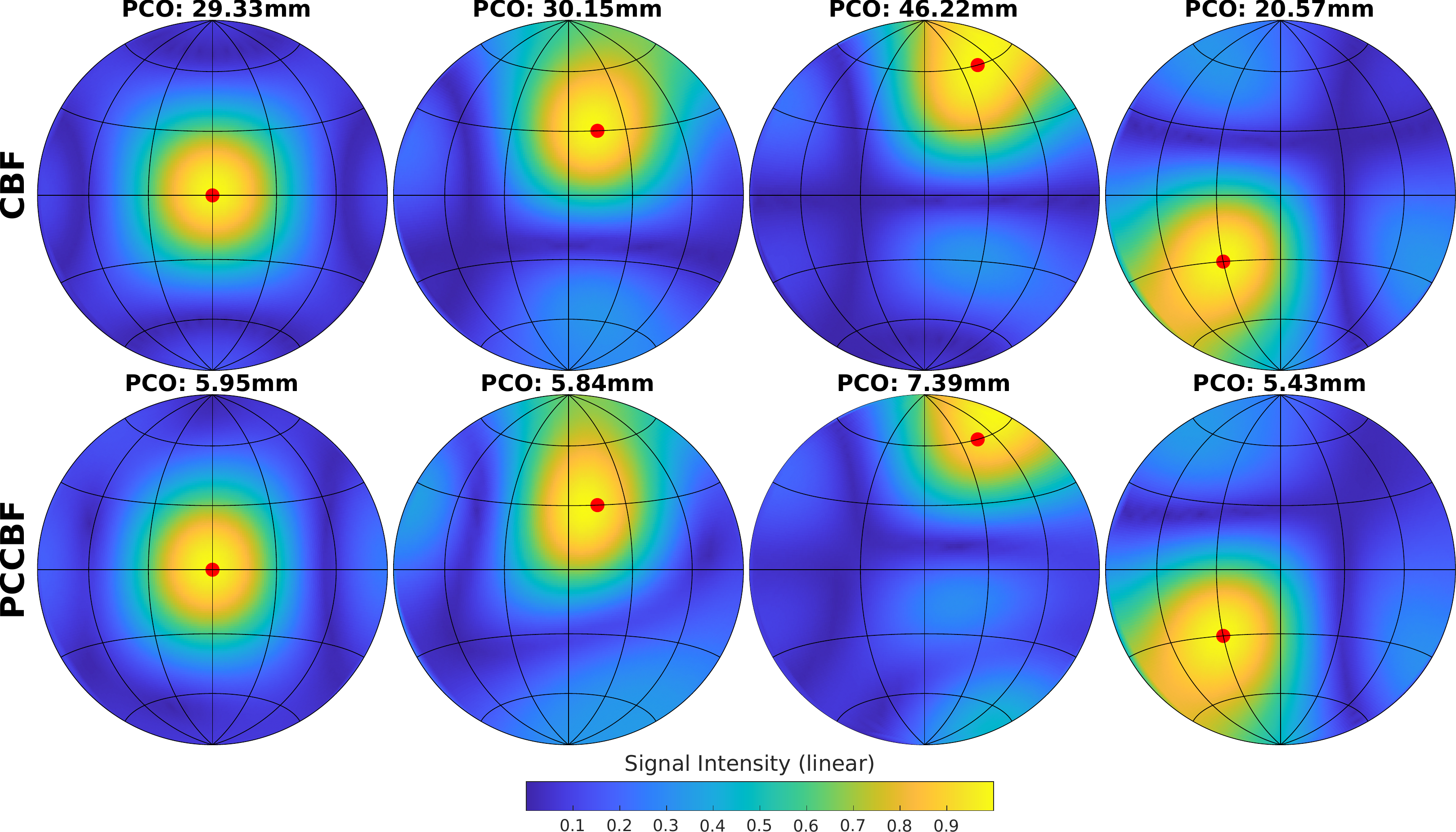}
    \caption{Comparison of beamforming results between Conventional Beamforming (CBF) and the proposed Phase-Center-Constrained Beamforming (PCCB) for four selected steering directions (indicated by red dots). The directivity patterns are shown on a linear scale using a Lambert Azimuthal Equal Area projection. The corresponding Phase Center Offset (PCO) norm is indicated above each plot, demonstrating a significant reduction in PCO for PCCB while preserving directional gain.} 
    \label{fig:PCCB_vs_CBF}
\end{figure} 

To solve the PCCB algorithm, we cast the following optimization problem:
\begin{equation}
\min_w \|\vec{P}_c(w)\|_2^2 \quad \text{s.t.} \quad w^H\vec{v}(\psi_d)=1
\end{equation}
In this problem, we try to minimize the $\mathcal{L}_2$-norm under the constraint that the gain of the beamformer in the direction $\psi_d$ is equal to 1. We can rewrite this problem using earlier definitions, yielding:
\begin{equation}
\min_w \|D^{-1} \cdot  \Phi(w)\|_2^2
\end{equation}
in which we can replace $D^{-1}$ with $\Omega$ for notational consistency:

\begin{equation}
\min_w \|\Omega \cdot \Phi(w)\|_2^2
\end{equation}
Using the matrix expansion of the squared $\mathcal{L}_2$-norm, we get the following expression:
\begin{equation}
\min_w \Phi^T \cdot \Omega^T \cdot \Omega \cdot \Phi 
\end{equation}
which can be reduced for notational brevity by using $S= \Omega^T \cdot \Omega$ into:
\begin{equation}
\min_w \Phi^T \cdot S \cdot \Phi 
\label{eq:prob_final}
\end{equation}

In principle, we can combine equation \ref{eq:prob_final} with the constraint $w^H\vec{v}(\psi_d)=1$, and solve this using a constrained Sequential Quadratic Programming (SQP) approach. However, when one would implement this approach, the risk exists that the beampattern $B_w(\psi)$ shows excessive gain into other directions than the desired look direction $\psi_d$, as the constraint $w^H\vec{v}(\psi_d)=1$ only constraints the gain to be equal to 1 in direction $\psi_d$, and does not constrain the pattern in any other way. To overcome this, we take inspiration from the approach followed in the Minimum Variance Distortionless Response (MVDR) beamformer, where the overall signal intensity is minimized. In what follows, we detail these regularization techniques, and expand the linear constraints to accommodate nulling into directions of jammers or interfering sources.

\textbf{Linear constraints:} We expand the original constraint $w^H\vec{v}(\psi_d)=1$ into a constraint matrix $C$ and a response vector $g$ as follows:
\begin{equation}
w^H C = g,
\end{equation}
in which we can express gain and (first-order) nulling constraints as follows:
\begin{equation}
C = [\vec{v}(\psi_d) \quad \vec{v}(\psi_n)], \quad g = [1 \quad 0]^T
\end{equation}
to implement a gain of 1 into direction $\psi_d$ and a null in direction $\psi_n$.

 \textbf{Regularization through energy regularization:} to avoid the excessive gains of the beampattern in unwanted directions, we minimize the overall energy of beampattern $B_w(\psi)$ by having a regularization term representing the squared $\mathcal{L}_2$-norm of the beampattern:
\begin{equation}
J_e = \frac{1}{N} w^H V V^H w
\end{equation}
which penalizes excessive energies in $B_w(\psi)$. Note that we add in a scaling factor $\frac{1}{N}$ in which $N$ is the number of sensor elements. 

\textbf{Original Beampattern Fidelity:} The final regularization term used is the least-squares difference of the normalized beampattern made with the steering vector of the desired directions $\psi_d$ and the resulting beampattern for weights $w$=
\begin{equation}
J_b =  \left\| \frac{B_d(\psi)}{ \left\|B_d(\psi) \right\|} - \frac{B_w(\psi)}{ \left\|B_w(\psi) \right\|} \right\|^2_2
\end{equation}
in which we can define the desired beampattern $B_d(\psi)$ as:
\begin{equation}
B_d = \frac{1}{L}\sum_{l=1}^L \vec{v}(\psi_{d,j})\cdot V
\end{equation}
for the $L$ directions of interest to which we set a gain of 1 in $g$. Including these constraints and fidelity terms, we now cast the final functional J as follows:
\begin{equation}
J_t(w) = \Phi^T \cdot S \cdot \Phi + \lambda_e J_e + \lambda_b J_b
\end{equation}
with $\lambda_e$ and $\lambda_b$ being parameters deciding on the degree of regularization. We can then plug this equation into our final optimization problem:
\begin{equation}
w^* = \arg\min_w J_t(w), \quad \text{s.t.} \quad w^H \cdot C = g
\end{equation}
This problem can be solved using constrained sequential quadratic programming. However, many implementations do not work well with complex constraint functions. Therefore, we expand the constraint terms by splitting the real and imaginary parts:

\begin{figure*}[t]
    \centering
    \includegraphics[width=\linewidth]{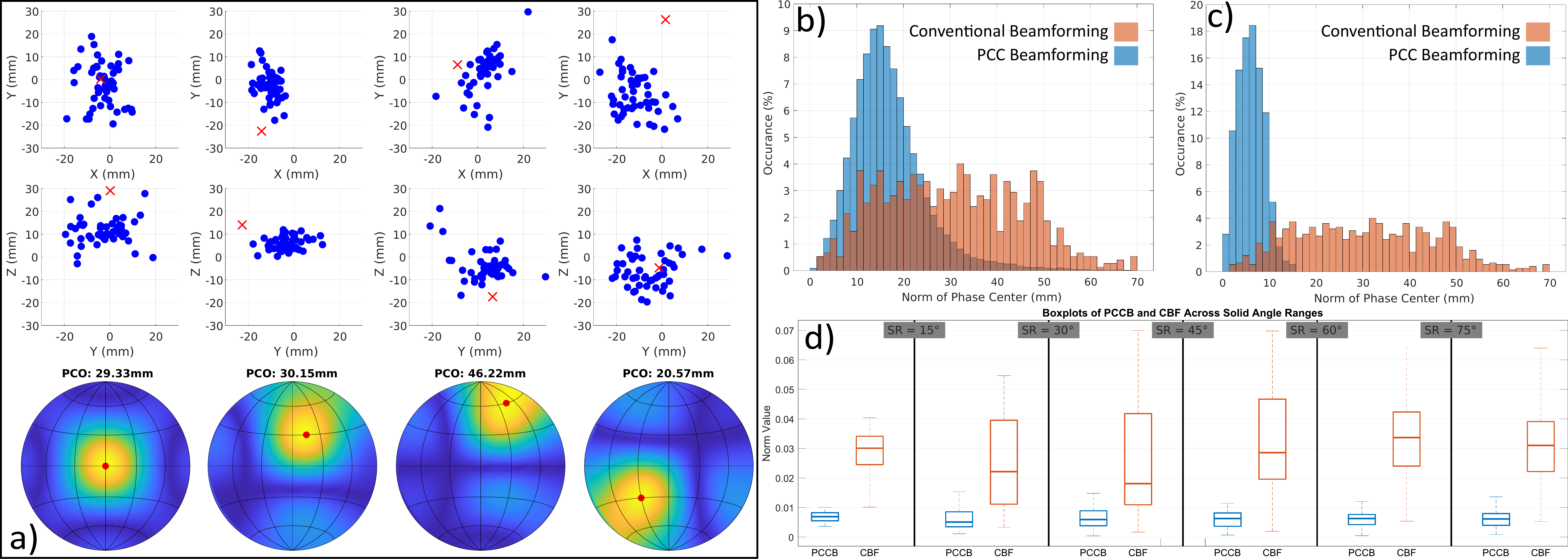}
    \caption{Statistical analysis of the Phase-Center-Constrained Beamforming (PCCB) algorithm across multiple random initializations and steering directions. (a) shows the projection of the Phase Center Offset (PCO) locations onto the XY and YZ planes. Blue dots represent the PCOs obtained from PCCB across 50 random initializations per steering direction, while red crosses indicate the PCOs resulting from Conventional Beamforming (CBF). Directivity patterns for selected steering directions are shown below the corresponding projections. (b) shows the histogram of the PCO norms across all 50 runs for each steering direction, comparing PCCB (blue) with CBF (orange). (c) shows the histogram of the PCO norms for the best (i.e., lowest PCO norm) solution among the 50 random initializations for each steering direction, again comparing PCCB (blue) with CBF (orange). Finally, (d) shows the boxplots of PCO norms as a function of the steering range (SR), which groups steering directions by their angular distance from boresight. The plot shows that PCCB maintains a consistently low PCO norm across the entire steering range, with no significant degradation at larger steering angles.} 
    \label{fig:statistics}
\end{figure*} 

\begin{equation}
\Gamma_r = \text{Re}\{w^H C - g\}, \quad \Gamma_i = \text{Im}\{w^H C - g\}
\end{equation}
Stacking these constraints yields:
\begin{equation}
\Gamma = \begin{bmatrix}\Gamma_r & \Gamma_i\end{bmatrix}^T 
\end{equation}

Thus, the final form of the optimization is expressed as:
\begin{equation}
w^* = \arg\min_w J_t(w), \quad \text{s.t.} \quad \Gamma=\vec{0}
\label{eq:finalfinal}
\end{equation}
This is the problem we use for the remainder of the paper. We set $\lambda_e=10$ and $\lambda_b=0.1$, as this yielded the best results in our experimentation.

\section{Experimental Validation}
To validate the proposed PCCB algorithm, we simulated a GNSS antenna array of 9 elements, divided onto a square grid with spacing of 7cm in both dimensions (see figure \ref{fig:coord}). We simulated the system on the L1 frequency band of GNSS (1575.42MHz), and used a standard patch antenna model for the elements. We performed a set of experiments in which we chose a desired steering direction $\psi_d$, and calculated the weights using optimization problem \ref{eq:finalfinal}. We solved the problem using the Matlab 2024a implementation of constrained SQP, with random initializations for the solution. We repeated the random initialization 50 times, and solved the problem until convergence with a step size tolerance of $1\cdot 10^{-9}$. Using the obtained weights as well as the weights obtained from the steering vector into the desired direction $\psi_d$, we calculated the PCO using equation \ref{eq:PCO}. Using these raw measurements, we performed several analyses, which we will detail in the following subsections.

\subsection{Beamforming results versus PCO norm}
As a first validation of the implemented PCCB approach, we show the resulting beampatterns of the PCCB approach and compare it to the beampatterns obtained using Conventional Beamforming (CBF). In figure \ref{fig:PCCB_vs_CBF} we show the beampatterns for four chosen directions $\psi_d$, indicated by the red dot in the figure. The plots are on a linear scale, and shown using a Lambert Azimuthal Equal Area projection. Above each of the plots, we indicate the norm of the PCO coordinate (which is a point in 3D space, see figure \ref{fig:coord}). From this plot, it becomes clear that the norm of the PCO is reduced by a factor 5, while still retaining comparable spatial filtering to the spatial filtering obtained from the CBF approach.

\subsection{Analysis of PCCB Stability}
The proposed PCCB approach is a numerical non-linear optimization problem, involving complex constraints. These problems are inherently non-convex, and can therefore suffer from local minima, and excessive dependence on the initial conditions of the solution. As stated before, we ran the optimization algorithm 50 times for each steering direction $\psi_d$, to analyze the variability of the found solution. A first glimpse into the statistics of the PCCB algorithm can be seen in figure \ref{fig:statistics}, panel a). Here, we show the projection of the PCO onto the XY and YZ planes, both for the CBF approach (red cross) as well as for the PCCB approach (blue dots). A significant spread can be seen in the found PCO coordinates, indicating the potential of the presence of local minima in the optimization functional $J_t$. The directivity patterns of chosen directivity $\psi_d$ are shown below the projections for reference. Panels b) and c) show the histograms of the PCO norms, both for the conventional beamforming (orange) as well as for the PCCB approach (blue). Panel b) shows the PCO norms for all runs and steering directions, panel c) shows the histogram of the PCO norms where the 'best' solution of the 50 runs for each steering direction was chosen. In both cases, the PCO norm obtained using PCCB is significantly lower compared to the norms obtained from CBF, and the norm where the 'best' solution was chosen is even further reduced compared to the complete run. This indicates the usefulness of random initialization of the algorithm, and solving the SQP problem multiple times. Finally, panel d) shows the boxplot of the PCO norm for various regions of the directivity pattern, indicated by the opening angle SR. No significant difference in PCO norm can be observed for increasing steering angles more towards the periphery, which is especially interesting for GNSS applications where satellites can occupy the full sky above the antennas.

\section{Conclusion and Future work}
In this paper we introduced a novel approach to beamforming which takes into account the location of the phase center of the complete antenna beampattern, which we called Phase-Center-Constrained Beamforming (PCCB). The PCCB algorithm is designed to minimize the Phase Center Offset (PCO) displacement in antenna arrays, which is particularly relevant for high-precision GNSS applications. Experimental validations demonstrated that PCCB significantly reduces the PCO compared to conventional beamforming methods, achieving approximately a fivefold improvement while maintaining comparable spatial filtering performance. Stability analyses highlighted the algorithm's susceptibility to local minima, caused by the nonlinear cost function that is being optimized. This susceptibility emphasizes the benefit of multiple randomized initializations to achieve consistently optimal solutions. Importantly, PCCB showed robust performance across various steering directions, including peripheral angles, which is particularly valuable in GNSS scenarios where uniform sky coverage is essential. 

In future research, we will focus on further improving the numerical stability of the algorithm, and investigate more deeply the possible constraints that can be provided to the algorithm. Furthermore, more advanced solving methods for the SQP problem should be investigated which are less susceptible to the local minimal in the cost surface.
\section*{Disclaimer }
The content of the present article reflects solely the authors’ view and by no means represents the official ESA view.
\bibliographystyle{IEEEtran}
\bibliography{references_Zotero}
\end{document}